# Kadomtsev-Petviashvili (KP) Burgers equation in dusty negative ion plasmas: Evolution of dust-ion acoustic shocks

A. N. Dev[1], J. Sarmah[2], M. K. Deka[3], A. P. Misra[4] and N. C. Adhikary[5, #]

[1]Department of Science and Humanities, College of Science and Technology, RUB, Bhutan
[2]Department of Mathematics, R. G. Baruah College, Guwahati-781025, Assam
[3]Centre of Plasma Physics, Tepesia, Sonapur, Kamrup, Assam, India
[4]Department of Mathematics, Siksha Bhavana, Visva-Bharati University, Santiniketan-731 235, India.
[5]Physical Sciences Division, Institute of Advanced Study in Science and Technology, Vigyan Path, Paschim Boragaon, Garchuk, Guwahati – 781035, Assam, India.

# Corresponding authors e-mail: nirab_iasst@yahoo.co.in; Ph. No: +91 99540-77133

**Abstract:** We study the nonlinear propagation of dust-ion acoustic (DIA) solitary waves in an unmagnetized dusty plasma which consists of electrons, both positive and negative ions and negatively charged immobile dust grains. Starting from a set of hydrodynamic equations with the ion thermal pressures and ion kinematic viscosities included, and using a standard reductive perturbation method, the Kadomtsev-Petviashivili Burgers (KPB) equation is derived, which governs the evolution of DIA shocks. A stationary solution of the KPB equation is obtained and its properties are analysed with different plasma number densities, ion temperatures and masses. It is shown that a transition from shocks with negative potential to positive one occurs depending on the negative ion concentration in the plasma and the obliqueness of propagation of DIA waves.

## 1. Introduction

In the last few decades, numerous investigations have been made on the study of nonlinear waves and structures in low-temperature dusty plasmas as the presence of extremely massive charged dust particles plays a vital role in understanding the electrostatic disturbances in space plasma environments [1-3] as well as laboratory plasma devices [4,5]. It has been pointed out that the static charged dust grains can drastically modify the existing response of electrostatic wave spectra in dusty plasmas [4–11]. On the other hand, depending upon whether the dust grains are static or mobile, there appear new types of electrostatic waves including solitary or shock waves in plasmas. Shukla and Silin first theoretically [12] investigated the existence of low-frequency dust-ion acoustic (DIA) waves (with phase speed much smaller than the electron thermal speed and larger than the ion thermal speed) in a three-component dusty plasma. Later, this DIA wave was experimentally observed by Barkan *et al* using a dusty plasma device [13].

The Korteweg-de Vries (KdV) equation, which governs the evolution of solitary waves, was first derived by Washimi and Tanuiti in a normal two-component plasma [14] using a reductive perturbation technique. It has been reported that multi-component plasmas in presence of sufficient amount of negative ions can support both compressive and rarefactive KdV solitons [15]. Furthermore, the properties of DIA solitary waves and shocks as observed in laboratory dusty plasmas are well explained by the modified KdV and KdV–Burgers (KdVB) equations [5,11,16,17].

In the novel work of Shukla [17], the DIA shocks and holes were described by the KdVB equation in presence of ion kinematic viscosity in dusty plasmas. In a similar work, Rahman *et al* [18] studied the characteristics of dust-acoustic (DA) shocks in a adiabatic dusty plasma. It has been confirmed that the presence of negative ions in dusty plasmas also plays an important role in many aspects including charging of dust particles. Such negative ions in plasmas can also lead to the formation of rarefactive DIA solitons [5].

It is well known that in a nonlinear dispersive media shock waves are formed due to an interplay between the nonlinearity (causing wave steepening) and dissipation (e.g., caused by viscosity, collisions, wave particle interaction, etc.) [19-22]. However, when a medium exhibits both dispersive and dissipative properties, the propagation characteristics of small-amplitude perturbations can be adequately described by KdVB (in one-dimension) or KPB (in two or three dimensios) equations. Many powerful methods have been established and developed to study these nonlinear equations by using standard reductive perturbation method, but one-dimensional form of expression cannot explain the complete picture of the solitary waves formed in nature. In 1970, Kadomtsev and Petviashvilli [23] proposed a multi-dimensional dispersive wave equation to study the stability of one soliton solution of the KdV equation under the influence of weak transverse perturbations. This KP equation is a partial differential equation to describe the nonlinear wave motion in more than one dimension and till then the KP equation has been considerably used in describing the nonlinear plasma dynamics.

In a theoretical study, Duan [24] had considered transverse perturbations and studied the propagation of DA solitary waves in an un-magnetized plasma in the framework of KP equation. Gill *et al.* [25] analyzed the properties of KP solitons in plasmas with two temperature ions and reported the formation of both the rarefactive and compressive solitons. In a similar study, Masood *et al* [26] reported the formation of two-dimensional nonplanar electrostatic shocks in an un-magnetized asymmetric pair-ion plasma and found that the kinematic viscosity enhances the shock strength. In an another work, Dorranian *et al.* [27], derived a KP equation using reductive perturbation method and studied the effects of nonthermal ions on the solitons in a dusty plasma. They also showed that the formation of compressive and rarefactive solitary waves are strongly dependent on the number density and temperature of nonthermal ions. Furthermore, KP equations for nonlinear DIA waves in dusty plasmas with warm ions as well as in pair-ion plasmas were investigated [28, 29].

In this work, we investigate the nonlinear propagation of DIA waves in a dusty plasma with a pair of ions. Using the standard reductive perturbation technique, a KPB equation is derived which governs the dynamics of DIA shocks. The effects of distinct ion temperatures, plasma number densities as well as the kinematic viscosities of ion fluids on the properties of DIA shocks are discussed.

## 2. Theoretical formulation

We consider the propagation of DIA waves in a multi-component dusty plasma consisting of inertialess electrons, a pair of singly charged inertial ions and immobile negatively charged dust particles. At equilibrium, the overall charge neutrality condition reads $n_{e0} = n_{p0} - n_{n0} - n_{d0}z_d$, where $n_{j0}$ is the equilibrium number density of j-th species particle, $z_{d0}$ is the number of electrons residing on the dust grain surface and *e* is the elementary charge. The dynamics of electrons and ions can be expressed by the following set of hydrodynamic equations:

$$\frac{\partial n_j}{\partial t} + \nabla \cdot \left( n_j \vec{v}_j \right) = 0 \qquad \text{----- (1)}$$

$$\left(\frac{\partial}{\partial t}+\vec{v}_j.\nabla\right)\vec{v}_j = \mp\frac{Z_j q_j}{m_j}\nabla\varphi - \frac{\nabla\vec{P}_j}{m_j n_j} + \mu_j \nabla^2 \vec{v}_j \qquad \text{----- (2)}$$

where $n_j$, $v_j$, and $m_j$, respectively, denote the number density, fluid velocity, and mass of $j$-species ions with $q_{p(n)} = e(-e)$ for positive (negative) ions. In Eq. (2), we have used the adiabatic equation of state for ions, i.e., $P_j/P_{j0} = (n_j/n_{j0})^\gamma = N_j^\gamma$, where $P_{j0} = n_{j0}\kappa_B T_j$, $\gamma = \frac{5}{3}$ for three dimensional geometry of the system and $T_j$ is the particle's thermodynamic temperature. The inertialess electron density is given by the Boltzmann distribution

$$n_e = n_{e0}\exp\left(\frac{e\varphi}{\kappa_B T}\right) \qquad \text{----- (3)}$$

Where $\varphi$ is the electrostatic potential and $\kappa_B$ is the Boltzmann constant and finally the Poisson's equation given by

$$\nabla^2\varphi = -4\pi e\left(n_p - Z_d n_d - n_n - n_e\right) \qquad \text{----- (4)}$$

We normalize the Eqs. (1)-(4) according to $\phi = \frac{e\varphi}{k_B T_e} \Leftrightarrow \varphi = \frac{k_B T_e}{e}\phi$, $N_j = \frac{n_j}{n_{j0}}$, $V_j = \frac{v_j}{C_s}$; where $C_s = \sqrt{\frac{\kappa_B T_e}{m_n}}$ is the DIA speed. The space and time variables are normalized by the Debye length $\lambda_D = \left(\frac{k_B T_e}{4\pi n_{n0} e^2}\right)^{1/2}$ and the inverse of the negative-ion plasma frequency, $\omega_{pn} = \left(\frac{4\pi n_{n0} e^2}{m_n}\right)^{1/2}$ respectively where $\omega_{pn}\lambda_d = C_s$. Furthermore, $\rho_j = \frac{\mu_j}{\omega_{pn}\lambda_D^2}$ is the non-dimensional ion kinematic viscosity.

Thus, Eqs. (1) to (3) can be recast as

$$\frac{\partial(N_p)}{\partial \tau} + \nabla.\left(N_p \vec{V}_P\right) = 0 \qquad \text{----- (5)}$$

$$\frac{\partial(N_n)}{\partial \tau} + \nabla.\left(N_n \vec{V}_n\right) = 0 \qquad \text{----- (6)}$$

$$\left(\frac{\partial}{\partial \tau}+\vec{V}_p.\nabla\right)\vec{V}_p = -m\nabla\phi - \frac{5}{3}\sigma_p m \frac{\nabla N_P}{N_P^{1/3}} + \rho_p \nabla^2\left(\vec{V}_p\right) \qquad \text{----- (7)}$$

$$\left(\frac{\partial}{\partial \tau}+\vec{V}_n.\nabla\right)\vec{V}_n = \nabla\phi - \frac{5}{3}\sigma_n \frac{\nabla N_n}{N_n^{1/3}} + \rho_n \nabla^2\left(\vec{V}_n\right) \qquad \text{----- (8)}$$

$$\nabla^2\phi = \left(\mu_e \exp(\phi) + N_n - \mu_p N_p - \mu_d\right) \qquad \text{----- (9)}$$

where $\mu_e = \frac{n_{e0}}{n_{n_0}}$, $\mu_p = \frac{n_{p0}}{n_{n0}}$, $\mu_d = \frac{n_{d0}z_d}{n_{n0}}$, $m = \frac{m_n}{m_p}$, $\sigma_p = \frac{T_p}{T_e}$, $\sigma_n = \frac{T_n}{T_e}$

## 3. Derivation of KP-Burgers Equation

In order to derive the KPB equation from Eqs. (4) to (9) we use the reductive perturbation technique with the stretched coordinates:

$$\xi = \varepsilon^{\frac{1}{2}}(x - \lambda t), \quad \eta = \varepsilon y, \quad \varsigma = \varepsilon z, \quad \tau = \varepsilon^{\frac{3}{2}} t, \quad \rho_j = \varepsilon^{\frac{1}{2}} \rho_{j0}$$

where $\varepsilon$ is a smallness parameter measuring the weakness of the perturbation and $\lambda$ is the wave phase speed (normalized by $C_s$). The dependent physical variables, namely $n_d$, $u_d$, $z_d$ and $\phi$ are expanded in powers of $\varepsilon$ as

$$\left.\begin{aligned}
N_j &= 1 + \varepsilon N_j^{(1)} + \varepsilon^2 N_j^{(2)} + \varepsilon^3 N_j^{(3)} + \dots \\
V_{jx} &= \varepsilon V_{jx}^{(1)} + \varepsilon^2 V_{jx}^{(2)} + \varepsilon^3 V_{jx}^{(3)} + \dots \\
V_{jy,z} &= \varepsilon^{\frac{3}{2}} V_{jy,z}^{(1)} + \varepsilon^{\frac{5}{2}} V_{jy,z}^{(2)} + \varepsilon^{\frac{7}{2}} V_{jy,z}^{(3)} + \dots \\
\phi &= \varepsilon \phi^{(1)} + \varepsilon^2 \phi^{(2)} + \varepsilon^3 \phi^{(3)} + \dots
\end{aligned}\right\} \quad \text{----- (10)}$$

In what follows, we substitute the stretched coordinates and the expansions (10) into Eqs. (5) to (9), and equate the coefficients of different powers of $\varepsilon$.

In the lowest order of $\varepsilon$, i.e., $\left(\varepsilon^{3/2}\right)$ we get the first-order quantities:

$$N_p^{(1)} = \frac{m}{\left(\lambda^2 - \frac{5}{3} m\sigma_p\right)} \phi^{(1)} \Rightarrow N_p^{(1)} = \alpha_1 \phi^{(1)} \quad \text{----- (11)}$$

$$V_p^{(1)} = \frac{\lambda m}{\left(\lambda^2 - \frac{5}{3} m\sigma_p\right)} \phi^{(1)} \Rightarrow V_p^{(1)} = \alpha_1 \lambda \phi^{(1)} \quad \text{----- (12)}$$

$$N_n^{(1)} = \frac{-\phi^{(1)}}{\left(\lambda^2 - \frac{5}{3}\sigma_n\right)} \Rightarrow N_n^{(1)} = -\alpha_2 \phi^{(1)} \quad \text{----- (13)}$$

$$V_n^{(1)} = \frac{-\phi^{(1)}\lambda}{\left(\lambda^2 - \frac{5}{3}\sigma_n\right)} \Rightarrow V_p^{(1)} = -\alpha_2 \lambda \phi^{(1)} \quad \text{----- (14)}$$

together with the dispersion relation

$$\lambda^2 = \frac{1}{2\mu_e}\left[S + \sqrt{S^2 - \frac{20}{3}\mu_e m\left\{(\sigma_p + \sigma_n \mu_p) + \frac{5}{3}\sigma_p \sigma_n \mu_e\right\}}\right] \quad \text{----- (15)}$$

with $\alpha_1 = \dfrac{m}{\left(\lambda^2 - \frac{5}{3}m\sigma_p\right)}$, $\alpha_2 = \dfrac{1}{\left(\lambda^2 - \frac{5}{3}\sigma_n\right)}$ and $S = \left(\frac{5}{3}\mu_e m\sigma_p + 1\right) + \left(\frac{5}{3}\mu_e \sigma_n + \mu_p m\right)$

From the coefficients of the next higher-order of $\varepsilon$, i.e., $\left(\varepsilon^{5/2}\right)$ we have

$$\frac{\partial N_p^{(1)}}{\partial \tau} - \lambda \frac{\partial N_p^{(2)}}{\partial \xi} + \frac{\partial V_p^{(2)}}{\partial \xi} + \frac{\partial \left(N_p^{(1)} V_p^{(1)}\right)}{\partial \xi} + \frac{\partial V_p^{(1)}}{\partial \eta} + \frac{\partial V_p^{(1)}}{\partial \zeta} = 0 \quad \text{-----(16)}$$

$$\frac{\partial N_n^{(1)}}{\partial \tau} - \lambda \frac{\partial N_n^{(2)}}{\partial \xi} + \frac{\partial V_n^{(2)}}{\partial \xi} + \frac{\partial \left(N_n^{(1)} V_n^{(1)}\right)}{\partial \xi} + \frac{\partial V_n^{(1)}}{\partial \eta} + \frac{\partial V_n^{(1)}}{\partial \zeta} = 0 \quad \text{-----(17)}$$

$$\frac{\partial V_p^{(1)}}{\partial \tau} - \lambda \frac{\partial V_p^{(2)}}{\partial \xi} - \frac{1}{3}\lambda N_p^{(1)} \frac{\partial V_p^{(1)}}{\partial \xi} + V_p^{(1)} \frac{\partial V_p^{(1)}}{\partial \xi} =$$
$$-m\frac{\partial \phi^{(2)}}{\partial \xi} - m\frac{1}{3} N_p^{(1)} \frac{\partial \phi^{(1)}}{\partial \xi} - \frac{5}{3}m\sigma_p \frac{\partial N_p^{(2)}}{\partial \xi} + \rho_{p0} \frac{\partial^2 V_{px}^{(1)}}{\partial \xi^2} \quad \text{-----(18)}$$

$$\frac{\partial V_n^{(1)}}{\partial \tau} - \lambda \frac{\partial V_n^{(2)}}{\partial \xi} - \frac{1}{3}\lambda N_n^{(1)} \frac{\partial V_n^{(1)}}{\partial \xi} + V_n^{(1)} \frac{\partial V_n^{(1)}}{\partial \xi} =$$
$$\frac{\partial \phi^{(2)}}{\partial \xi} + \frac{1}{3} N_n^{(1)} \frac{\partial \phi^{(1)}}{\partial \xi} - \frac{5}{3}\sigma_n \frac{\partial N_n^{(2)}}{\partial \xi} + \rho_{n0} \frac{\partial^2 V_n^{(1)}}{\partial \xi^2} \quad \text{-----(19)}$$

$$\frac{\partial}{\partial \xi}\left(\frac{\partial V_p^{(1)}}{\partial \eta}\right) + \frac{\partial}{\partial \xi}\left(\frac{\partial V_p^{(1)}}{\partial \zeta}\right) = \frac{m}{\lambda}\left(\frac{5}{3}\sigma_p \alpha_1 + 1\right)\left(\frac{\partial^2 \phi^{(1)}}{\partial \eta^2} + \frac{\partial^2 \phi^{(1)}}{\partial \zeta^2}\right) \quad \text{-----(20)}$$

$$\frac{\partial}{\partial \xi}\left(\frac{\partial V_n^{(1)}}{\partial \eta}\right) + \frac{\partial}{\partial \xi}\left(\frac{\partial V_n^{(1)}}{\partial \zeta}\right) = -\frac{1}{\lambda}\left(\frac{5}{3}\sigma_n \alpha_2 + 1\right)\left(\frac{\partial^2 \phi^{(1)}}{\partial \eta^2} + \frac{\partial^2 \phi^{(1)}}{\partial \zeta^2}\right) \quad \text{-----(21)}$$

and $\quad \dfrac{\partial^2 \phi^{(1)}}{\partial \xi^2} = \mu_e \phi^{(2)} + \dfrac{1}{2}\mu_e \left(\phi^{(1)}\right)^2 + N_n^{(2)} - \mu_p N_p^{(2)} \quad \text{-----(22)}$

Finally, substituting the values from Eqs. (11) to (15) into Eqs. (16) to (22) and eliminating the $N_n^{(2)}$, $N_p^{(2)}$, $V_n^{(2)}$, $N_p^{(2)}$ and $\varphi^{(2)}$ we obtain the required KPB equation as

$$\frac{\partial}{\partial \xi}\left[\frac{\partial \phi^{(1)}}{\partial \tau} + A\phi^{(1)}\frac{\partial \phi^{(1)}}{\partial \xi} + B\frac{\partial^3 \phi^{(1)}}{\partial \xi^3} - C\frac{\partial^2 \phi^{(1)}}{\partial \xi^2}\right] + D\left(\frac{\partial^2 \phi^{(1)}}{\partial \eta^2} + \frac{\partial^2 \phi^{(1)}}{\partial \zeta^2}\right) = 0 \quad \text{-----(23)}$$

where the coefficients of nonlinearity and dispersion are, respectively, A and B. The coefficients C and D appear due to the ion kinematic viscosities and the transverse perturbations respectively. These are given by

$$A = \left[\frac{\mu_p\left(3\lambda^2 - 5\sigma_n\right)\left(8\lambda^2 \alpha_1^2 + m\alpha_1\right) - \left(3\lambda^2 - 5m\sigma_p\right)\left\{\left(8\lambda^2 \alpha_2^2 + \alpha_2\right) - 3\mu_e\left(3\lambda^2 - 5\sigma_n\right)\right\}}{6\left\{\alpha_2 \lambda\left(3\lambda^2 - 5m\sigma_p\right) + \alpha_1 \lambda \mu_p\left(3\lambda^2 - 5\sigma_n\right)\right\}}\right]$$

$$B = \left[\frac{\left(3\lambda^2 - 5\sigma_n\right)\left(3\lambda^2 - 5m\sigma_p\right)}{2\left\{\alpha_2 \lambda\left(3\lambda^2 - 5m\sigma_p\right) + \alpha_1 \lambda \mu_p\left(3\lambda^2 - 5\sigma_n\right)\right\}}\right],$$

$$C = \left[\frac{\alpha_2 \rho_{n0} \lambda (3\lambda^2 - 5m\sigma_p) + \mu_p \rho_{p0} \alpha_1 \lambda (3\lambda^2 - 5\sigma_n)}{2\{\alpha_2 \lambda (3\lambda^2 - 5m\sigma_p) + \alpha_1 \lambda \mu_p (3\lambda^2 - 5\sigma_n)\}}\right],$$

$$D = \left[\frac{(5\sigma_n \alpha_2 + 3)(3\lambda^2 - 5m\sigma_p) + \mu_p m(5\sigma_p \alpha_1 + 3)(3\lambda^2 - 5\sigma_n)}{2\{\alpha_2 \lambda (3\lambda^2 - 5m\sigma_p) + \alpha_1 \lambda \mu_p (3\lambda^2 - 5\sigma_n)\}}\right].$$

Equation (23) governs the evolution of small-amplitude DIA shocks in dusty pair-ion plasmas. A travelling wave solution of the KPB equation (23) can readily be obtained. To this end we use the transformation $\chi = (l\xi + m\eta + n\zeta - U\tau)$, where $l$, $m$, $n$ are the direction cosines along the $x$, $y$, $z$ axes and let $\phi^{(1)}(\xi,\eta,\zeta,\tau) = \psi(\chi)$. Thus, from Eq. (23) one obtains

$$Bl^3 \frac{d^2\psi}{d\chi^2} - Cl^2 \frac{d\psi}{d\chi} + Al\frac{\psi^2}{2} - U\psi + D(m^2 + n^2) = 0 \quad \text{----- (24)}$$

We now apply the *tanh*-method in which we define $z = \tanh(\chi), \psi(\chi) = w(z)$. Then Eq. (24) becomes

$$(1-z^2)^2 l^4 \frac{d^2 w}{dz^2} - \left(2l^4 z + \frac{C}{B}l^2\right)(1-z^2)\frac{dw}{dz} + \frac{Al^2}{2B^2}w^2 - (Dl^2 + Ul - D)w = 0 \quad \text{----- (25)}$$

For the series solution of Eq. (25) we assume $w(z) = \sum_{r=0}^{\infty} a_r z^{\delta+r}$ and then the leading order analysis of finite terms gives $r = 2$ and $\delta = 0$ so that $w(z)$ becomes

$$w(z) = a_0 + a_1 z + a_2 z^2$$

Now, substituting the value of $w(z)$ in Eq. (25), one can obtain the values of $a_0$, $a_1$, $a_2$ as

$$a_0 = \frac{\{(Dl^2 + Ul - D) + 12Bl^4\}}{Al^4}, \quad a_1 = -\frac{12Cl}{5B}, \quad a_2 = -\frac{12Bl^2}{A}$$

Thus, a required solution of Eq. (23) is given by

$$\phi^{(1)} = \frac{(Dl^2 + Ul - D)}{Al^2} - \frac{12Cl}{5A}\tanh(\chi) + \frac{12Bl^2}{A}\text{sech}^2(\chi) \quad \text{----- (26)}$$

where $C = 10Bl$ and $(Dl^2 + Ul - D) = 24Bl^4$.

## 4. Results and discussion

In this section we numerically examine the nonlinear wave speed $\lambda$, the coefficients of the KPB equation as well as shock profile (26) with a set of parameters that are representative of laboratory and space plasmas. Typical plasma parameters are considered as $m_{p0}=m_{n0}=40\times1.6^{-27}$kg, $n_{e0}=3.8\times10^{14}$m$^{-3}$, $n_{p0}=5.4\times10^{14}$ m$^{-3}$, $n_{n0}=0.3\times10^{14}$ m$^{-3}$, $n_{d0}=1.2\times10^{10}$ m$^{-3}$, $z_{d0}=1.5\times10^4$e, $T_e=1$ eV, $T_p=T_n=0.1$ eV [5, 30].

In Fig. 1 the nonlinear wave speed $\lambda$ [Here we consider only the fast wave corresponding to the '+' sign in Eq. (15), as in most laboratory situations slow waves (corresponding to the '-' sign) do not favour the formation of solitary waves or shocks. Also, with the negative sign, $\lambda$ varies as $S/\mu_e$, *i.e.*, phase speed $\lambda$ remains almost constant with the density ratio] is plotted against the density ratio $\mu_e$ for different values of $m$, $\sigma_p$ (~ $\sigma_n$) and $\mu_d$. It is found that the value of $\lambda$ decreases with increasing values of $\mu_e$ and always larger than the DIA speed. It is also found that the ion mass ratio 'm' has an important effect on the wave speed $\lambda$. As the mass ratio increases an enhancement of the wave speed is seen to occur. However, the normalised dust density $\mu_d$ reduces the wave speed for the same plasma parameters.

Next, we investigate interplay among the coefficients of the KPB equation (23) with various dusty plasma parameters. In the sub figure 2(a) the variation of the nonlinear coefficient A with the electron to negative ion density ratio ($\mu_e$) is shown for different values of $m$, $\sigma_p$ and $\mu_d$. The solid, dashed, dotted and dash-dotted lines, respectively, correspond to $m = 1$, $\sigma_p = \sigma_n = 0.07$, $\mu_d = 6$; $m = 1.1$, $\mu_p = \mu_n = 0.07$, $\mu_d = 6$; $m = 1$, $\sigma_p = 0.5$; $\sigma_n = 0.07$, $\mu_d = 6$ and $m = 1$, $\sigma_p = \sigma_n = 0.07$, $\mu_d = 5$. It is clear that the nonlinear coefficient *A* always increases with $\mu_e$ and also with increasing the ion mass ratio; but decreases when $\mu_d$ is reduced (dash-dot line). The sub figure 2(b) shows the variation of the dispersive coefficient B with $\mu_e$ for different values of $m$, $\sigma_p$ and $\mu_d$. This shows that the dispersive coefficient *B* is seen to be decreased with $\mu_e$ and also with decreasing $\mu_d$ (dash-dot line); but *B* increases with increasing the ion mass ratio *m* (dashed line). For $\rho_{p0} = 0.2$ and $\rho_{n0} = 0.01$ the dissipative coefficient C also increases with $\mu_e$ and *m*; but decreases as $\mu_d$ reduces (dash-dot line), [See Fig. 2(c)]. As shown in the sub figure 2(d), the transverse coefficient D decrease with $\mu_e$ and $\mu_d$ but increases with increasing values of *m*.

Figure 3 shows the profile of the DIA shocks for different plasma parameters. The (a)left and (b) right panels exhibit profiles with negative and positive potentials for a fixed value of the direction cosine $l = 0.3$ and other plasma parameters are as $m = 1$, $\sigma_p = \sigma_n = 0.07$, $\mu_d = 6$ and $U = 10$. From the figure, it is seen that the transition from shocks with negative potential to positive one occurs at $\mu_e \geq 2.2$. However, for a fixed $\mu_e = 2.2$ and others as $m = 1$, $\sigma_p = \sigma_n = 0.07$, $\mu_d = 6$ and $U = 10$, the transition from shocks with positive potential to negative one occurs at $l < 0.3$. In the left panel (a), the solid line (red) shows the shock profile corresponds to the plasma parameters $m = 1$, $\sigma_p = \sigma_n = 0.07$, $\mu_d = 6$; $\mu_e = 1.6$. The shock amplitude decreases for reduction of both $\mu_e$ (indicated by the dashed line) and $\mu_d$ (thick line). On the other hand, shock amplitude increases for increasing both $\sigma_p$ (dotted line) and *m* (dash-dotted line). However, as shown in the right panel (b), the solid line shows the shock profile correspond to the parameter values $\mu_e = 3$, $m = 1$, $\sigma_p = \sigma_n = 0.07$, $\mu_d = 6$. Here the shock amplitude increases with enhancement of $\mu_e$ (dashed line) and with reduction of $\mu_d$ (thick line). But shock amplitude decreases with enhancement of both $\sigma_p$ (dotted line) and *m* (dashed-dot line).

## 5. Conclusion

We have investigated the nonlinear propagation of DIA shocks in an un-magnetized dusty negative ion plasma. Using the standard reductive perturbation technique, a KP-Burgers equation is derived which governs the dynamics of small-amplitude DIA shocks in dusty pair-ion plasmas. It is shown that the different magnitudes of ion temperatures, mass and number density of both the ions, electron density, dust particle charge together play a crucial role in the formation of DIA shocks in multi-component dusty plasmas. It is also found that there is a critical density of electron to negative ion density ratio for which a transition of shocks with negative potential to positive one occurs. On the other hand, dust density, charge and ion temperature controls the amplitude of the shock profiles.

**References**
1. Horanyi M and Mendis D A 1985 *Astrophys J.* **294** 357.
2. Goertz C K 1989 *Rev. Geophys.* **27** 271.
3. Whipple E C , Northrop T G and Mendis D A 1985 *J. Geophys. Res.* **90** 7405.

4. Barkan A, D'Angelo N, Merlino R L 1994 *Phys. Rev. Lett.* **73** 3093.
5. Adhikary N C, Deka M K and Bailung H 2009 *Phys Plasmas* **16** 063701.
6. Misra A P, Chowdhury A R and Chowdhury K R 2004 *Phys. Lett. A* **323** 110.
7. Misra A P, Chowdhury K R and Chowdhury A R 2007 *Phys. Plasmas* **14** 012110.
8. Saini N S and Kourakis I 2008 *Phys. Plasmas* **15** 123701.
9. Mamun A A 2008 *Phys Lett. A* **372** 4610.
10. Mamun A A, Cairns R A and Shukla P K 2009 *Phys Lett. A* **373** 2355.
11. Gill T S, Bains A S, Saini N S and Bedi C 2010 *Phys Lett. A* **374** 3210.
12. Shukla P K and Silin V P 1992 *Phys. Scripta* **45** 508.
13. Barkan A, D'Angelo N, Merlino R L 1996 *Planet. Space Sci.* **44** 239.
14. Washimi H and Taniuti T 1966 *Phys. Rev. Lett.* **17** 996.
15. Das G C and S. G. Tagare S G 1975 *Plasma Phys.* **17** 1025.
16. Nakamura Y, Bailung H, Shukla P K 1999 *Phys. Rev. Lett.* **83** 1602.
17. Shukla P K 2000 *Phys. Plasmas* **7** 1044.
18. Rahman A, Sayed F and Mamun A A 2007 *Phys. Plasmas* **14** 034503.
19. Xue J K 2003 *Eur. Phy. J. D* **26** 211.
20. Adhikary N C 2012 *Phys. Lett. A* **376** 1460.
21. Misra A P, Adhikary N C and Shukla P K 2012 *Phys Rev. E* **86** 056406.
22. Mamun A A and M.S. Zobaer M S 2014 *Phys Plasmas* **21** 022101.
23. Kadomtsev B B and Petviashvili V I 1970 *Sov. Phys. Dokl.* **15** 539.
24. Duan W S 2001 *Phys. Plasmas* **8** 3583.
25. Gill T S, Saini N S and Harvinder K 2006 *Chaos Soliton. Fract.* **28** 1106.
26. Masood W and Rizvi H 2012 *Phys Plasmas* **19** 012119.
27. Dorranian D and Sabetkar A 2012 *Phys. Plasmas* **19** 013702.
28. El-Wakil S A, Abulwafa E M, El-Shewy E K, Gomaa H and Abd-El-Hamid H M 2013 *Astrophys Space Sci* **346** 141.
29. Rehman H U 2013 *Chin. Phys. B* **22** 035202
30. Adhikary N C, Bailung H, Pal A R, Chutia J and Nakamura Y 2007 *Phys Plasmas* **14** 103705.

**Figure Captions:**

**Figure 1**: The nonlinear wave speed $\lambda$ (fast wave, considering the '+' sign) is plotted against the density ratio $\mu_e$ for different values of $m$, $\sigma_p$ (~ $\sigma_n$) and $\mu_d$ as in the figure. The value of $\lambda$ is seen to decrease with increasing values of $\mu_e$ and always larger than the dust-acoustic speed.

**Figure 2**: The variations of the nonlinear and dispersive coefficients (a) A, (b) B, (c) C and (d) D with respect to $\mu_e$ are shown for different values of $m$, $\sigma_p$ and $\mu_d$. The values of B and D remain almost unaltered with $\sigma_p$.

In sub figure (a), the solid, dashed, dotted and dash-dotted lines, respectively, correspond to $m = 1$, $\sigma_p = \sigma_n = 0.07$, $\mu_d = 6$; $m = 1.1$, $\mu_p = \mu_n = 0.07$, $\mu_d = 6$; $m = 1$, $\sigma_p = 0.5$; $\sigma_n = 0.07$, $\mu_d = 6$ and $m = 1$, $\sigma_p = \sigma_n = 0.07$, $\mu_d = 5$.

In sub figure (b), the solid, dashed and dash-dotted lines, respectively, correspond to $m = 1$, $\sigma_p = \sigma_n = 0.07$, $\mu_d = 6$; $m = 2$, $\sigma_p = \sigma_n = 0.07$, $\mu_d = 6$ and $m = 1$, $\sigma_p = \sigma_n = 0.07$, $\mu_d = 2$.

In sub figure (c), the solid, dashed, dotted and dash-dotted lines, respectively, correspond to $m = 1$, $\sigma_p = \sigma_n = 0.07$, $\mu_d = 6$; $m = 1.1$, $\sigma_p = \sigma_n = 0.07$, $\mu_d = 6$; $m = 1$, $\sigma_p = 0.5$, $\sigma_n = 0.07$, $\mu_d = 6$ and $m = 1$, $\sigma_p = \sigma_n = 0.07$, $\mu_d = 5$ with $\rho_{p0} = 0.2$ and $\rho_{n0} = 0.01$.

In sub figure (d), the solid, dashed and dash-dotted lines, respectively, correspond to $m = 1$, $\sigma_p = \sigma_n = 0.07$, $\mu_d = 6$; $m = 2$, $\sigma_p = \sigma_n = 0.07$, $\mu_d = 6$ and $m = 1$, $\sigma_p = \sigma_n = 0.07$, $\mu_d = 2$.

**Figure 3**: The profiles of the dust-acoustic shocks with (a) negative and (b) positive potentials are shown. For example, for a fixed $l = 0.3$ and others as $m = 1$, $\sigma_p = \sigma_n = 0.07$, $\mu_d = 6$ and $U = 10$, the transition from shocks with negative potential to positive one occurs at $\mu_e \geq 2.2$.

However, for a fixed $\mu_e = 2.2$ and others as $m = 1$, $\sigma_p = \sigma_n = 0.07$, $\mu_d = 6$ and $U = 10$, the transition from shocks with positive potential to negative one occurs at $l < 0.3$.

In the left panel (a), the solid, dashed, dotted, dash-dotted and the thick lines, respectively, correspond to the variations with $\mu_e$, $\sigma_p$, m and $\mu_d$ with the parameter values $\mu_e = 1.5$, $m = 1$, $\sigma_p = \sigma_n = 0.07$, $\mu_d = 6$; $\mu_e = 1.6$, $m = 1$, $\sigma_p = \sigma_n = 0.07$, $\mu_d = 6$; $\mu_e = 1.5$, $m = 1$, $\sigma_p = 0.2$; $\sigma_n = 0.07$, $\mu_d = 6$; $\mu_e = 1.5$, $m = 1.1$, $\sigma_p = \sigma_n = 0.07$, $\mu_d = 6$ and $\mu_e = 1.5$, $m = 1$, $\sigma_p = \sigma_n = 0.07$, $\mu_d = 5$. The other fixed values are $l = 0.3$ and $U = 10$.

In the right panel (b), the solid, dashed, dotted, dash-dotted and the thick lines, respectively, correspond to the variations with $\mu_e$, $\sigma_p$, m and $\mu_d$ with the parameter values $\mu_e = 3$, $m = 1$, $\sigma_p = \sigma_n = 0.07$, $\mu_d = 6$; $\mu_e = 3.2$, $m = 1$, $\sigma_p = \sigma_n = 0.07$, $\mu_d = 6$; $\mu_e = 3$, $m = 1$, $\sigma_p = 0.2$; $\sigma_n = 0.07$, $\mu_d = 6$; $\mu_e = 3$, $m = 1.02$, $\sigma_p = \sigma_n = 0.07$, $\mu_d = 6$ and $\mu_e = 3$, $m = 1$, $\sigma_p = \sigma_n = 0.07$, $\mu_d = 5$. The other fixed values are $l = 0.3$ and $U = 10$.

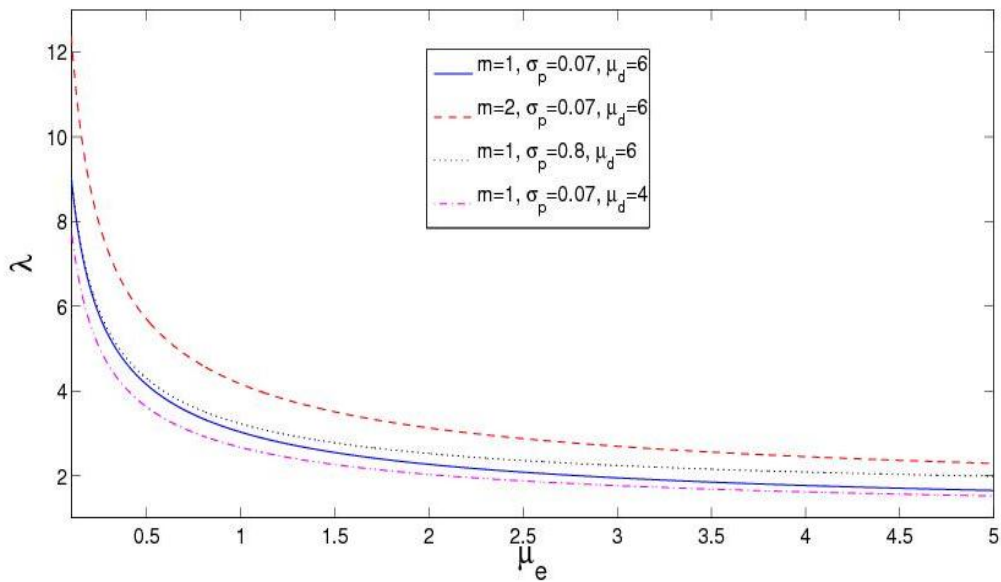

Fig.1

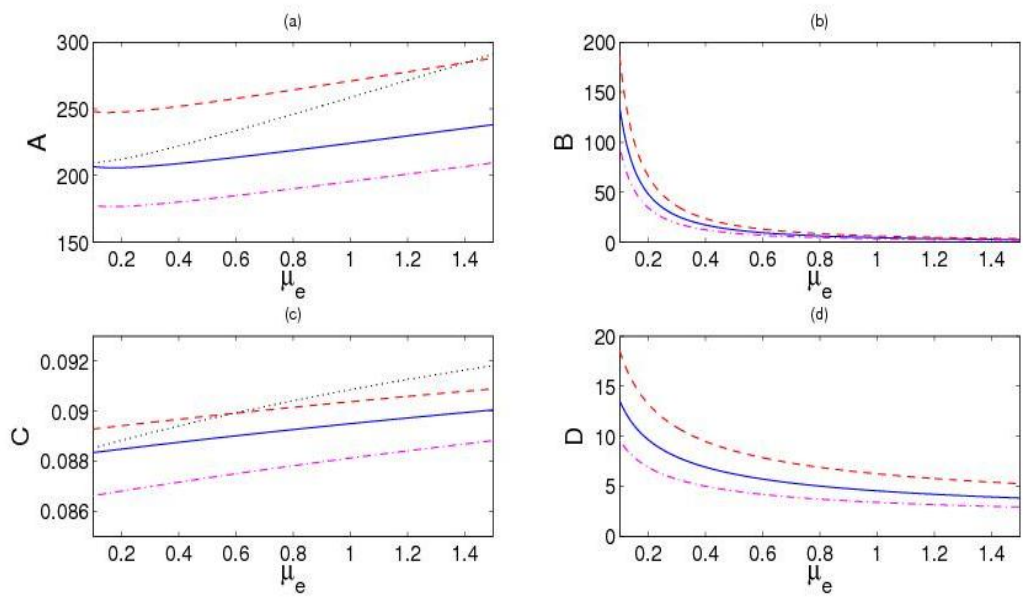

Fig.2

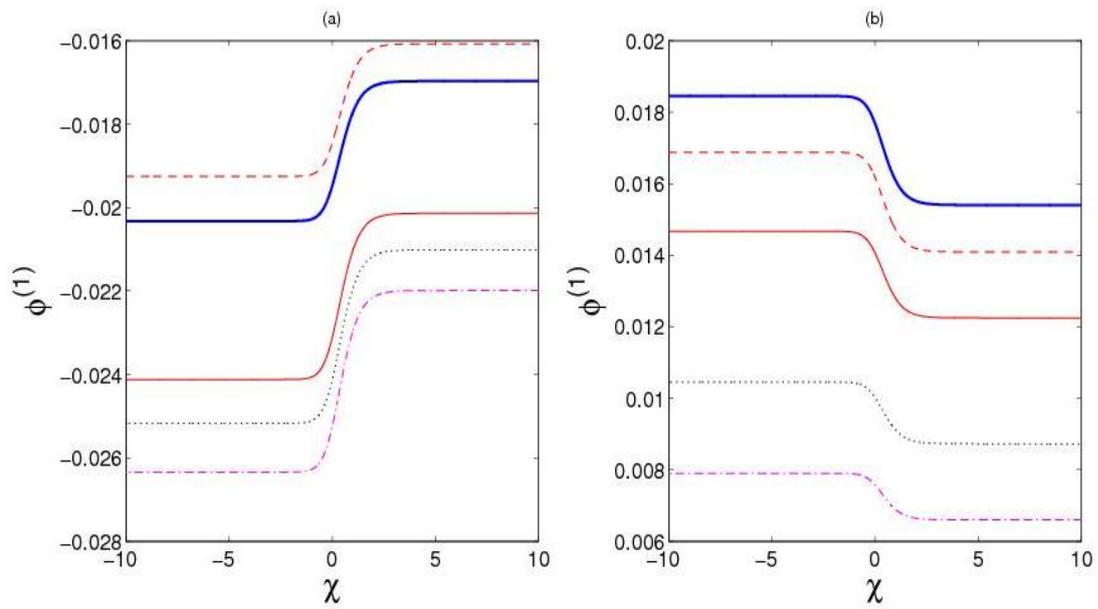

Fig.3